\input harvmac

\font\cmss=cmss10
\font\cmsss=cmss10 at 7pt
\def\IZ{\relax\ifmmode\mathchoice
{\hbox{\cmss Z\kern-.4em Z}}{\hbox{\cmss Z\kern-.4em Z}}
{\lower.9pt\hbox{\cmsss Z\kern-.4em Z}}
{\lower1.2pt\hbox{\cmsss Z\kern-.4em Z}}\else{\cmss Z\kern-.4em
Z}\fi}

\lref\BMN{
D.~Berenstein, J.~M.~Maldacena and H.~Nastase,
``Strings in flat space and pp waves from N = 4 super Yang Mills,''
JHEP {\bf 0204}, 013 (2002)
[arXiv:hep-th/0202021].
}

\lref\seven{
N.~R.~Constable, D.~Z.~Freedman, M.~Headrick, S.~Minwalla, L.~Motl,
A.~Postnikov and W.~Skiba,
``PP-wave string interactions from perturbative Yang-Mills theory,''
arXiv:hep-th/0205089.
}

\lref\SpradlinAR{
M.~Spradlin and A.~Volovich,
``Superstring interactions in a pp-wave background,''
arXiv:hep-th/0204146.
}

\lref\BerensteinSA{
D.~Berenstein and H.~Nastase,
``On lightcone string field theory from super Yang-Mills and holography,''
arXiv:hep-th/0205048.
}

\lref\fgm{
D. Freedman, U. Gursoy and L. Motl, in progress.
}

\lref\KiemXN{
Y.~Kiem, Y.~Kim, S.~Lee and J.~Park,
``pp-wave / Yang-Mills correspondence: An explicit check,''
arXiv:hep-th/0205279.
}

\lref\HuangWF{
M.~Huang,
``Three point functions of N=4 Super Yang Mills from light cone string
field theory in pp-wave,''
arXiv:hep-th/0205311.
}

\lref\GopakumarDQ{
R.~Gopakumar,
``String interactions in PP-waves,''
arXiv:hep-th/0205174.
}

\lref\MetsaevBJ{
R.~R.~Metsaev,
``Type IIB Green-Schwarz superstring in plane wave Ramond-Ramond  background,''
Nucl.\ Phys.\ B {\bf 625}, 70 (2002)
[arXiv:hep-th/0112044].
}

\lref\MetsaevRE{
R.~R.~Metsaev and A.~A.~Tseytlin,
``Exactly solvable model of superstring in plane wave
Ramond-Ramond  background,''
Phys.\ Rev.\ D {\bf 65}, 126004 (2002)
[arXiv:hep-th/0202109].
}

\lref\GSB{
M.~B.~Green, J.~H.~Schwarz and L.~Brink,
``Superfield Theory Of Type II Superstrings,''
Nucl.\ Phys.\ B {\bf 219}, 437 (1983).
}

\lref\GreenTC{
M.~B.~Green and J.~H.~Schwarz,
``Superstring Interactions,''
Nucl.\ Phys.\ B {\bf 218}, 43 (1983).
}

\lref\BergmanHV{
O.~Bergman, M.~R.~Gaberdiel and M.~B.~Green,
``D-brane interactions in type IIB plane-wave background,''
arXiv:hep-th/0205183.
}

\lref\DasguptaHX{
K.~Dasgupta, M.~M.~Sheikh-Jabbari and M.~Van Raamsdonk,
``Matrix perturbation theory for M-theory on a PP-wave,''
JHEP {\bf 0205}, 056 (2002)
[arXiv:hep-th/0205185].
}

\lref\GrossSU{
D.~J.~Gross, A.~Mikhailov and R.~Roiban,
``Operators with large R charge in N = 4 Yang-Mills theory,''
arXiv:hep-th/0205066.
}

\lref\KristjansenBB{
C.~Kristjansen, J.~Plefka, G.~W.~Semenoff and M.~Staudacher,
``A new double-scaling limit of N = 4 super Yang-Mills theory and PP-wave
strings,''
arXiv:hep-th/0205033.
}

\lref\ChuPD{
C.~S.~Chu, V.~V.~Khoze and G.~Travaglini,
``Three-point functions in N=4 Yang-Mills theory and pp-waves,''
arXiv:hep-th/0206005.
}

\lref\Herman{
H.~Verlinde,
``Bits, Matrices and 1/N,''
arXiv:hep-th/0206059.
}

\lref\basisone{
N.~Beisert, C.~Kristjansen, J.~Plefka, G.~W.~Semenoff and M.~Staudacher,
``BMN correlators and operator mixing in N = 4 super Yang-Mills theory,''
arXiv:hep-th/0208178.
}

\lref\basistwo{
D.~J.~Gross, A.~Mikhailov and R.~Roiban,
``A calculation of the plane wave string Hamiltonian from N = 4
super-Yang-Mills theory,''
arXiv:hep-th/0208231.
}

\lref\basisthree{
N.~R.~Constable, D.~Z.~Freedman, M.~Headrick and S.~Minwalla,
``Operator mixing and the BMN correspondence,''
arXiv:hep-th/0209002.
}

\lref\psvvv{
J.~Pearson, M.~Spradlin, D.~Vaman, H.~Verlinde and A.~Volovich,
``Tracing the String:  BMN Correspondence at Finite $J^2/N$,''
arXiv:hep-th/0210102.
}

\Title{\vbox{\baselineskip12pt
        \hbox{hep-th/0206073}
        \hbox{PUTP-2041}
        \hbox{HUTP-02/A023} 
}}{Superstring Interactions in a pp-wave Background II}

\centerline{
Marcus Spradlin${}^{1}$ and Anastasia Volovich${}^{2}$
}

\bigskip
\centerline{${}^{1}$~Department of Physics}
\centerline{Princeton University}
\centerline{Princeton, NJ 08544}
\centerline{\tt spradlin@feynman.princeton.edu}
\centerline{}
\centerline{${}^{2}$~Department of Physics}
\centerline{Harvard University}
\centerline{Cambridge, MA 02138}
\centerline{\tt nastya@gauss.harvard.edu}

\vskip .3in
\centerline{\bf Abstract}

In type IIB light-cone superstring field theory,
the cubic interaction has
two pieces:  a delta-functional overlap and an operator inserted
at the interaction point.  In this paper we extend
our earlier work hep-th/0204146 by computing the matrix elements of
this operator in the oscillator basis of pp-wave string theory
for all $\mu p^+ \alpha'$.
By evaluating these matrix elements for large $\mu p^+ \alpha'$, we
check a recent conjecture relating matrix elements
of the light-cone string field theory Hamiltonian
(with prefactor) to certain three-point
functions of BMN operators in the gauge theory.
We also make several explicit predictions for gauge theory.

\smallskip

\Date{}

\listtoc
\writetoc

\newsec{Introduction}

The proposal of \BMN\ equating string
theory in a pp-wave background
\refs{\MetsaevBJ,\MetsaevRE} to a certain
sector of ${\cal{N}}=4$ SU($N$) Yang-Mills theory containing
operators of large $R$-charge $J$
involves a limit in which the two parameters
\eqn\parameters{
\lambda' = {g_{\rm YM}^2 N \over J^2} = {1
\over (\mu p^+ \alpha')^2}, \qquad g_2 = {J^2 \over N}
= 4 \pi g_s (\mu p^+ \alpha')^2
}
are held fixed while $N$ and the 't Hooft coupling
$\lambda = g_{\rm YM}^2 N$ are taken
to infinity (see also \GrossSU).
It has recently been proposed \refs{\KristjansenBB \BerensteinSA-\seven}
to extend this duality to include
string interactions.
The existence of two expansion parameters \parameters\ 
opens up the intriguing possibility that there is
a regime in which both the string and gauge theories
are effectively perturbative.

The authors of \seven\ postulated
a correspondence between Yang-Mills perturbation theory and
string perturbation theory of the form
\eqn\conjecture{
\langle i | H | j\rangle |k \rangle =
\mu (\Delta_i - \Delta_j - \Delta_k) C_{ijk},
}
to leading order in $\lambda'$, and for a particular
class of states.  Here
$i, j, k$ label three free string states in the pp-wave
background and the left-hand side is the matrix element of
the interacting light-cone string field theory Hamiltonian which connects
the two-string state $|j\rangle |k\rangle$ to the single string
state $\langle i|$.
On the right-hand side, $C_{ijk}$ is the coefficient of the three-point
function for the operators corresponding to the three string states
and $\Delta_i$ is the conformal dimension of operator $i$ (equivalently,
the mass of string state $i$).
The conjecture \conjecture\ seemed to pass a nontrivial test in a check
of the one-loop mass renormalization of the  pp-wave
string spectrum
\seven.
Although the leading (in $g_2$) term in the light-cone
string field theory Hamiltonian was determined formally in \SpradlinAR\ for
all $\lambda'$, the precise matrix elements investigated
in \seven\ had not been checked explicitly in string
theory.

Type IIB light-cone superstring field theory for the
pp-wave was constructed in \SpradlinAR. 
The cubic interaction
has two parts:  a delta-functional overlap
which expresses continuity of the string worldsheet, and
the `prefactor', which is an operator
required by supersymmetry that is inserted at the
point where the string splits.
At $\lambda' = 0$ (infinite $\mu$), it is easy to
see \seven\ that
the rules for calculating the delta-functional overlap
agree with the rules for calculating the three-point functions of BMN
operators (see also \HuangWF, and \ChuPD\ for the leading
correction to this result).
Therefore
the nontrivial part of the proposal \conjecture\ is
that the effect of this prefactor (for large $\mu$) is
simply
to multiply the delta-functional overlap by the factor $\Delta_i
- \Delta_j - \Delta_k$.\foot{The fact that the
effective strength of string
interactions in the pp-wave background scales as
the energy of the string states involved
has been argued in \GopakumarDQ.}
In this paper we 
compute
the matrix elements of this
prefactor 
in the oscillator basis of pp-wave string theory for all $\lambda'$.
The leading $\lambda'$ behavior of the matrix elements we calculate
is not consistent with the proposal \conjecture.
This discrepancy in no way invalidates the BMN
correspondence, since matrix elements of the Hamiltonian are
not invariant quantities, but depend on a choice of basis.
That is, the {\it eigenvalues} of the interacting Hamiltonian, as computed
on both sides of the duality, should agree, but there is no reason
to expect the {\it matrix elements} of the left-hand side of \conjecture,
computed in the natural string field theory basis (of
single- and multi-string states), to agree with the
matrix elements of the right-hand side of \conjecture, computed
in the natural field theory basis (of single- and multi-trace operators).

The most significant difficulty in this enterprise is
the fact that the limits $\mu \to \infty$ (high curvatures in
spacetime) and $n \to \infty$ (short distances
on the worldsheet) do not commute.
Specifically, the prefactor arises entirely from a short-distance
effect on the string worldsheet and must therefore be determined
first at finite $\mu$, with the $\mu \to \infty$ limit
taken at the end.
This technical difficulty raises an important conceptual
question as well, since it would seem that the prefactor
cannot possibly be seen in perturbative gauge theory expanded
around $\mu = \infty$.
However, the unitarity check of \seven\ suggests otherwise,
since the prefactor can
apparently be obtained by `cutting' genus-one diagrams in the gauge
theory.
Clearly, a deeper understanding of how these two theories
are related is very desirable.

The plan of the paper is the following.
In section 2 we review the relevant gauge theory results
and give a brief overview of the three-string vertex.
In section 3 we determine the prefactor in the oscillator basis
for all $\mu$ and present a simple formula valid for large
$\mu.$ In section 4 we calculate the
leading ${\cal{O}}(\lambda')$ term in the matrix elements and
compare with the proposal \conjecture.

Related work includes
the matrix string theory approach to string interactions
in the pp-wave background \GopakumarDQ,
open string interactions between D-branes in a pp-wave
\BergmanHV,
and matrix models for M-theory on the maximally
supersymmetric pp-wave  \DasguptaHX.

After this work was completed, the paper \Herman\ appeared which
presents a string bit formalism for interacting strings in
the pp-wave background, expected to be valid to all orders in
$g_2$ and leading order in $\lambda'$.  This complements the
work of our paper, since we work at leading order
in $g_2$ but for all $\lambda'$.

\newsec{Overview}

In this section we summarize the relevant results
of \BMN, \seven\ and \SpradlinAR\ and write
the matrix elements
which will be calculated from string theory in section 4.

\subsec{Field Theory}

Following  \refs{\BMN,\seven} we consider
the chiral operators
\eqn\chiral{\eqalign{
O^J &= {1 \over \sqrt{J N^J}} \Tr(Z^J),\cr
O^J_\phi &= {1 \over \sqrt{N^{J+1}}} \Tr(\phi Z^J),\cr
O^J_\psi &= {1 \over \sqrt{N^{J+1}}} \Tr(\psi Z^J),
}}
as well as the operator
\eqn\almostchiral{
O^J_n = {1 \over \sqrt{J N^{J+2}}} \sum_{k=0}^J
e^{2 \pi i n k/J} \Tr(\phi Z^k \psi Z^{J-l}),
}
which is chiral for $n=0$ and ``almost chiral,'' in a
controllable sense, for $n/J \ll 1$.
Here $\phi$, $\psi$ and $Z$ are three
orthogonal
complex linear combinations of the six real scalar
fields of ${\cal{N}} = 4$ Yang-Mills theory.
Without loss of generality we can take
\eqn\aaa{
\phi = {X^1 + i X^2 \over \sqrt{2}}, \qquad
\psi = {X^3 + i X^4 \over \sqrt{2}}, \qquad
Z = {X^5 + i X^6 \over \sqrt{2}}.
}

Free string theory in the pp-wave background is exactly
solvable \MetsaevBJ.  The anomalous dimensions of
the operators \chiral\ and \almostchiral\ were computed
in \BMN, leading to the following identification between
these BMN
operators and string states:\foot{We use $\alpha$
to denote the oscillators which were
called $a$ in \BMN\ and \seven.  A different
basis, related for $n \ne 0$ by $\alpha_n = {1 \over \sqrt{2}} (a_{|n|}
- i~{\rm sign}(n) a_{-|n|})$ and $\alpha_0 = a_0$, appears in \SpradlinAR\ and
will be used later in this paper.}
\eqn\dictionary{\eqalign{
O^J &\leftrightarrow |0; p^+\rangle,\cr
O_\phi^J &\leftrightarrow
\alpha_0^{\phi \dagger} |0;p^+\rangle,\cr
O_\psi^J &\leftrightarrow
\alpha_0^{\psi \dagger} |0;p^+\rangle,\cr
O_n^J &\leftrightarrow
\alpha_n^{\phi \dagger}
\alpha_{-n}^{\psi \dagger} |0;p^+\rangle,
}}
where $\alpha^\phi =
{1 \over \sqrt{2}} (\alpha^1 - i \alpha^2)$,
$\alpha^\psi = {1 \over
\sqrt{2}} (\alpha^3 -
i \alpha^4)$, and $J$ and $p^+$ are related by
\eqn\aaa{
J = 
\sqrt{\lambda} \mu p^+ \alpha'.
}
The dictionary \dictionary\ is precise only at
lowest order in $\lambda'$ and $g_2$.  At higher orders some of the
operators receive
wavefunction renormalizations and
mix with each other (as well as with multi-trace operators)
and the set of BMN operators
must be diagonalized in order to preserve the correspondence
with single string states.
We will use the notation
$|O^J\rangle$, $|O_\phi^J\rangle$, etc. to refer
to the states in \dictionary, without any corrections.

Using the proposal \conjecture, the matrix
elements
\eqn\want{\eqalign{
\langle O^{J}_n | H |O^{J_1}_m \rangle |O^{J_2}\rangle
&= {4 g_s \mu \over \pi}
(1-y)
{(n y + m) \over (n y - m)}  \sin^2(\pi n y),\cr
\langle O_n^{J} | H | O_\phi^{J_1} \rangle | O_\psi^{J_2} \rangle
&=
{4 g_s \mu \over \pi}
\sqrt{y(1-y)}
\sin^2(\pi n y),
}}
were obtained in \seven\ 
from field theory calculations\foot{We
use the relativistic normalization $\langle i|j\rangle
= p^+_{i} \delta(p^+_{i} - p^+_{j})$, and
the matrix elements \want\ both contain a factor
of $p^+ \delta(p^+ - p^+_{1} - p^+_{2})$
which we suppress.}.
Here $J=J_1 + J_2$ and $y=J_1/J$.
The goal of this paper is to check whether these matrix elements
can be obtained from light-cone string field theory using the proposal
\conjecture.

\subsec{String Theory}

The cubic term in the light-cone string field theory
Hamiltonian in the pp-wave background was
determined in \SpradlinAR.  As explained in the introduction,
it consists of two pieces.  The first is the delta-functional
overlap, which can be expressed as a state $|V\rangle$ in the
three-string Hilbert space as
\eqn\vertex{
|V\rangle = \exp \left[ {1 \over 2} \sum_{r,s=1}^3
\sum_{m,n=-\infty}^\infty a^{I \dagger}_{m(r)}
\overline{N}^{(rs)}_{mn} a^{J \dagger}_{n(s)} \delta_{IJ} \right] |0\rangle.
}
Here $r,s \in \{1,2,3\}$
label the three strings and $I,J \in \{1,\ldots,8\}$ label
the transverse directions.
An expression for the Neumann matrix elements $\overline{N}^{(rs)}_{mn}$
was given in \SpradlinAR.
We have omitted the fermionic
terms which will make a brief but significant appearance below.
The second piece is the prefactor which we
represent as an operator
$\widehat{H}$
acting on the delta-functional\foot{The overall
normalization was not determined in \SpradlinAR.  In \seven\ it
was argued that the effective string coupling for the class
of states under consideration is $g_s \mu$, and we employ
this result here.}
$|H\rangle = 2 \pi g_s \mu \widehat{H} |V\rangle.$  It is given by
\eqn\aaa{
\widehat{H} = K^I \widetilde{K}^I v_{IJ}(\Lambda),
}
where the operators $K$, $\widetilde{K}$ and $\Lambda$ are 
linear in string creation operators ($\Lambda$ being fermionic), and
$v^{IJ}$ is the tensor
\eqn\aaa{
\eqalign{
v^{IJ}&= \delta^{IJ} - {i \over \alpha} \gamma^{IJ}_{ab}
\Lambda^a \Lambda^b + {1 \over 6 \alpha^2}
\gamma^{IK}_{ab} \gamma^{JK}_{cd} \Lambda^a \Lambda^b \Lambda^c
\Lambda^d\cr
&~- {4 i\over 6! \alpha^3} \gamma^{IJ}_{ab}
\epsilon_{abcdefgh}\Lambda^c\Lambda^d\Lambda^e \Lambda^f
\Lambda^g \Lambda^h
+ {16 \over 8! \alpha^4} \delta^{IJ}
\epsilon_{abcdefgh}
\Lambda^a \Lambda^b \Lambda^c\Lambda^d\Lambda^e \Lambda^f
\Lambda^g \Lambda^h.
}}
Here $\Lambda$ is a positive chirality SO(8) spinor,
$a,b,\ldots$ are spinor indices, and
we use the notation $\alpha_{(r)} = \alpha' p^+_{(r)}$, with
$\alpha \equiv\alpha_{(1)} \alpha_{(2)} \alpha_{(3)}$.
Formal expressions
for the operators $K$, $\widetilde{K}$ and
$\Lambda$
were presented
in \SpradlinAR.
In order to compare with \want\ we will need to calculate
the matrix elements of $K$ and $\widetilde{K}$ in the number basis of
pp-wave string theory.

Since none of the states of interest in \want\ have fermionic
excitations (i.e., they are all primary string
states\foot{We emphasize that the operator \almostchiral\ is
not a chiral primary in the {\it field} theory, but the state it
corresponds to in the {\it string} theory is primary, i.e.,
the
lightest member of a supermultiplet.  Whenever we say
`primary' we mean a string state with no fermionic excitations, not
(necessarily) a supergravity state, which would correspond
to a chiral primary operator in the field theory.})
we will not need to know the matrix elements of $\Lambda$, and
indeed we can set all fermionic non-zero modes 
in $\Lambda$ to zero, so that
$\Lambda = \alpha_{(1)} \lambda_{(2)} - \alpha_{(2)} \lambda_{(1)}$
in terms of the fermionic zero modes $\lambda_{(r)}$.
The supermultiplet structure in the pp-wave background
was studied in detail in \MetsaevRE, and it was shown that
primaries live in the
$\lambda_{\rm R}^4$ component of the type IIB superfield.
(Recall that the SO(8) transverse symmetry of the pp-wave background is broken
to SO(4)$\times$SO(4)
by a mass term for the fermions proportional
to $\Pi = \Gamma^1 \Gamma^2 \Gamma^3 \Gamma^4$, and we are
using the notation $\lambda_{\rm R} = \ha (1 + \Pi) \lambda$.)
A simple counting
of fermionic zero modes
shows that only the quartic term in $v^{IJ}$ 
has the right number of fermions to
contribute to the matrix element of primary states, so for
our purposes we need only
\eqn\aaa{
v^{IJ} = {1 \over 6 \alpha^2} \gamma^{IK}_{ab} \gamma^{JK}_{cd}
\Lambda^a \Lambda^b \Lambda^c \Lambda^d.
}

It was pointed out
in\foot{Also L. Motl, private
communication.} \refs{\KiemXN} that the tensor structure of this
term has a remarkable
property when the spinor indices $a,b,c,d$
are restricted to the same
SO(4) subgroup (which will be the case for primaries):
\eqn\kiemetal{
\gamma^{i K}_{[ab} \gamma^{jK}_{cd]} = \delta^{ij}
\epsilon_{abcd}, \qquad
\gamma^{i' K}_{[ab} \gamma^{j'K}_{cd]} = -\delta^{i'j'}
\epsilon_{abcd}, \qquad
\gamma^{i K}_{[ab} \gamma^{j'K}_{cd]} = 0,
}
where $i,j \in \{1,2,3,4\}$ and $i',j' \in \{5,6,7,8\}$.
Recalling that string oscillations in the directions 1,2,3,4
correspond in the Yang-Mills theory to the insertion of $\phi$ and
$\psi$ impurities into a string of $Z$'s, while string oscillations
in the directions 5,6,7,8 correspond to the insertion of $D_\mu Z$
impurities, the formula \kiemetal\ gives a simple
prediction: the matrix elements analogous to \want\ should change
only by a sign when the $\phi$ and $\psi$ are replaced by two $D_\mu Z$'s,
while they should vanish for matrix elements
with one $\phi$ or $\psi$
and one $D_\mu Z$.
This prediction holds
not just for the supergravity modes considered
in \KiemXN, but indeed for all  primary string states.
It would be interesting to see how the $\IZ_2$ symmetry between
the two SO(4)'s arises in the field theory \fgm, where
no symmetry between the two SO(4)'s seems manifest.

\newsec{The Prefactor in the Oscillator Basis}

The  bosonic operators $K$, $\widetilde{K}$ 
which appear in the prefactor are
\eqn\kpm{
K = K_+ - K_-, \qquad \widetilde{K} = K_+ + K_-,
}
where $K_\pm$ may be defined by
their action on $|V\rangle$ by
the formulas
\eqn\kdef{\eqalign{
K_+ | V \rangle &=
-2 \pi \sqrt{-\alpha}
\lim_{\sigma \to \pi \alpha_{(1)}}
(\pi \alpha_{(1)} - \sigma)^{1/2}
( P_{(1)}(\sigma) + P_{(1)}(-\sigma)) |V\rangle,\cr
K_- | V\rangle &= - 2 \pi \sqrt{-\alpha}
\lim_{\sigma \to \pi \alpha_{(1)}}
(\pi \alpha_{(1)} - \sigma)^{1/2}
{1 \over 4 \pi} (\partial_\sigma X_{(1)}(\sigma)
+ \partial_\sigma X_{(1)}(-\sigma))|V\rangle
}}
(see \SpradlinAR\ for details).
Note that we suppress the transverse SO(8) index
$I$ throughout this
section, since it plays no role.
Using the expansions of $P$ and $X$ in terms of string
modes \SpradlinAR, it is easy to see that $K_+$ will be linear
in oscillators $a^\dagger_m$
with non-negative indices $m$, and $K_-$ will
be linear in oscillators with negative indices, i.e.,
they will be of the form
\eqn\kpmexpan{
K_+ = \sum_{r=1}^3  F^0_{(r)} a_{0(r)}^\dagger
+ \sum_{r=1}^3
\sum_{m=1}^\infty F^+_{m (r)} a_{m(r)}^\dagger,
\qquad
K_- = \sum_{m=1}^\infty F^-_{m(r)} a_{-m(r)}^\dagger,
}
where the components $F$ are to be determined.
We  use  the notation $F_{n(r)}$ to collect
$F^-$, $F^0$ and $F^+$ into a single vector with
index $n$ from $-\infty$ to $+\infty$, so that
\eqn\kkdef{
K = \sum_{r=1}^3 \sum_{m=-\infty}^\infty
F_{m(r)} a_{m(r)}^\dagger.
}

Obtaining the matrix elements $F_{n(r)}$ from the definition
\kdef\ is tricky,
so we will use an
alternate approach.  The operator $K$
may also be determined by enforcing invariance under
the pp-wave superalgebra (see
\SpradlinAR\ for details).  In particular, it must satisfy
\eqn\constraints{\eqalign{
\left[ \sum_{r=1}^3 P_{(r)}(\sigma), K \right]
= \left[ \sum_{r=1}^3 e_{(r)} X_{(r)}(\sigma), K \right] = 0,
}}
where $e_{(r)} = {\rm sign}(p^+_{(r)})$,
which we take to be $+1$ for $r=1,2$ (incoming strings) and $-1$ for
$r=3$ (outgoing string).
Substituting \kkdef\ into \constraints\ and using the mode
expansions from \SpradlinAR, we find the equations
\eqn\fequations{
\sum_{r=1}^3  X^{(r)} C_{(r)}^{1/2} F_{(r)} = \sum_{r=1}^3
\alpha_{(r)} X^{(r)} C_{(r)}^{-1/2} F_{(r)} = 0.
}
The matrices $X^{(r)}$ and $C_{(r)}$, as well as all other
matrices which appear later in this section, are cataloged
in appendix A.
From the form of the overlap
matrix $X^{(r)}$ it is clear
that the negative and non-negative components of \fequations\ decouple.
Thus there are two linearly independent solutions to
\constraints, which is good since we need precisely the two linear
combinations \kpm\ of the negative and non-negative modes.

The zero-component of the vector equation \fequations\ is immediately solved 
to give
\eqn\fzero{
F^0_{(1)} = - \sqrt{\mu \alpha_{(1)}} \alpha_{(2)}, \qquad
F^0_{(2)} =   \sqrt{\mu \alpha_{(2)}} \alpha_{(1)}, \qquad
F^0_{(3)} = 0.
}
The overall normalization, which is not fixed by
\constraints, is determined by comparing to the
prefactor for the three superparticle vertex obtained
in \SpradlinAR.
Now we proceed to solve for the positive and negative index
parts of $F$.

\subsec{Positive Indices}

For positive indices we obtain upon substituting \fzero\ the equations
\eqn\eqone{\eqalign{
\sum_{r=1}^3 A^{(r)} C^{-1/2} C_{(r)}^{1/2} F^+_{(r)} &= {1 \over \sqrt{2}}
\mu \alpha B,\cr
\sum_{r=1}^3 \alpha_{(r)} A^{(r)} C^{-1/2} C_{(r)}^{-1/2} F^+_{(r)} &=
{1 \over \sqrt{2}} \alpha B.
}}
By taking an appropriate linear 
combination of 
\eqone\ to eliminate the right-hand side, we find
\eqn\eqtwo{
\eqalign{
\sum_{r=1}^3 A^{(r)} C^{1/2} C_{(r)}^{-1/2} U_{(r)}
F^+_{(r)} = 0.
}}
Using the first identity in (A.4) to solve \eqtwo\ gives
\eqn\fsolone{
F^+_{(r)} = {1 \over \alpha_{(r)}}
C_{(r)}^{1/2} C^{1/2} U^{-1}_{(r)} A^{(r) {\rm T}} V,
}
where $V$ is an arbitrary vector.
Plugging \fsolone\ into the second equation determines
$V$ and hence
the complete solution
\eqn\fplus{
F^+_{(r)} = {1 \over \sqrt{2}} {\alpha \over \alpha_{(r)}}
C_{(r)}^{1/2} C^{1/2} U^{-1}_{(r)} A^{(r) {\rm T}} \Upsilon^{-1} B, \qquad
\Upsilon \equiv \sum_{r=1}^3 A^{(r)} U^{-1}_{(r)} A^{(r) {\rm T}}.
}
It is easy to see that
one recovers the flat space expression of \GSB\ upon
setting $\mu$ to zero
(modulo one factor of $C^{1/2}$ which
arises from a different normalization of
the string oscillators).

\subsec{Negative Indices}

For negative indices the corresponding equations
are
\eqn\eqtwo{\eqalign{
\sum_{r=1}^3 {1 \over \alpha_{(r)}}
A^{(r)} C^{1/2} C_{(r)}^{1/2} F_{(r)}^- &= 0,\cr
\sum_{r=1}^3 A^{(r)} C^{1/2} C_{(r)}^{-1/2} F_{(r)}^- &= 0.
}}
These equations are more subtle than the ones from the previous subsection.
This is because (as in flat space \GSB),
it turns out that the solution
for $F^-$ is such that the expressions \eqtwo\ actually diverge
when $F^-$ is substituted.  However, when one goes back to the function
of $\sigma$ responsible for the divergence, one finds that it is
of the form $\delta(\sigma - \pi \alpha_{(1)}) - \delta(\sigma +
\pi \alpha_{(1)})$, which we are allowed to ignore, since those two
points are identified (see \GSB\ for details).

Instead of solving \eqtwo\ directly,
we can bypass this subtlety by appealing to a trick.
In flat space, the trick one can appeal to is a very simple relation
between positive-index and negative-index Neumann matrix elements.
In flat space that relation is a consequence of conformal invariance
on the worldsheet,
which we do not have in light-cone gauge for pp-wave string theory.
Nevertheless we will be able to obtain a fairly simple relation which
suffices.

Recall the definition \kdef\ of $K_\pm$.
Using the expansion of $P(\sigma)$ and $X(\sigma)$
in terms of modes, and using \vertex, one finds that
\eqn\yyy{
F_{n(r)} \sim \lim_{\sigma \to \pi
\alpha_{(1)}} (\pi \alpha_{(1)}-
\sigma)^{1/2}\sum_{p=-\infty}^\infty 
s(p)
\sqrt{|p|} \cos {p \sigma \over \alpha_{(1)}}
\overline{N}^{(1r)}_{pn},
}
where $s(p) = 1$ for $p \ge 0$ and $s(p) = -i$ for $p < 0$.
The $\epsilon^{-1/2}$ singularity near
$\sigma = \pi \alpha_{(1)} - \epsilon$
comes entirely from the large $p$ behavior of this
sum.
Therefore, we can determine a relation between
$F^+$ and $F^-$ by determining a relation between
Neumann matrix elements for positive and negative indices.
For example, the relation
\eqn\zzz{
\left[ \overline{N}^{(s3)} \right]_{-p,-n} = -
\left[ U_{(s)} \overline{N}^{(s3)} U_{(3)} \right]_{pn}, \qquad
s \in \{1,2\}
}
is proven in appendix B, and other necessary relations are
easily proven for other components.
Note that the factor $U_{(s)}$ on the left
goes to $1$ for large $p$, so it does not affect the
large $p$ behavior.
We obtain from \zzz\ and the
analogous relations for other $r,s$ the remarkably simple
relation
\eqn\fminus{
F^-_{(r)} = i U_{(r)} F^+_{(r)}.
}

\subsec{Large $\mu$ Expansion}

The expressions
\fzero, \fplus\ and \fminus\ determine the matrix elements
of the prefactor for all $\mu$.
They are more explicit than
the formulas
\kdef\ which were presented in \SpradlinAR\ in
the continuum basis, but
we are still one step away from being able to precisely
compare
our matrix elements
to \want: we need to expand these expressions to leading
order for large $\mu$.
Unfortunately the formula \fplus\ is not well-suited for this
purpose.  One can show that
\eqn\uib{
\Upsilon^{-1} B \sim 
C B + {\cal{O}}(\mu^{-1})
}
up to an overall ($\mu$-independent) coefficient.
The leading term is annihilated by $A^{(r) {\rm T}}$ for
$r \in \{1,2\}$ and hence gives no contribution to \fplus, while
the subleading term has the property that multiplying
on the left by $A^{(r) {\rm T}}$ gives a divergent sum.
This problem alerts us to the fact that for
$r \in \{1,2\}$ the large $\mu$ expansion of \uib\ does
not commute with multiplication by $A^{(r) {\rm T}}$,
so one has to first calculate the entire matrix $A^{(r) {\rm T}}
\Upsilon^{-1} B$ and then take the limit.
Only for the $r=3$ component is it sensible and consistent
to use \uib, which leads to
\eqn\fthreelarge{
F^{(3)}_+ \sim - \mu^{-1/2} {1 \over 2\sqrt{2}} {\alpha \over
|\alpha_{(3)}|^{3/2}}
C^{5/2} B.
}

To find the leading behavior of the $r \in \{1,2 \}$ components
it is easier to
 return to the original equations \eqone, which for large
$\mu$ degenerate into
the single equation
\eqn\simple{
\sum_{r=1}^3 \sqrt{\mu \alpha_{(r)}}
A^{(r)} C^{-1/2} F_{(r)}^+ = {1 \over \sqrt{2}} \mu \alpha B,
}
whose solution\foot{One might be worried that a single equation
is not sufficient to determine two quantities $F^{(1)}$ and $F^{(2)}$.
However,  $A^{(1)}$
and $A^{(2)}$ have orthogonal images and no kernels, so the equation
$A^{(1)} v_{(1)} + A^{(2)} v_{(2)} = 0$ has no nontrivial solutions.
Hence the solution to \simple\ is indeed unique.} is
\eqn\frlarge{
F^+_{(r)} = - \sqrt{\mu \over 2} {\alpha \alpha_{(3)}
\over \alpha_{(r)}^{3/2}} C^{3/2} A^{(r) {\rm T}} C^{-1} B, \qquad
r \in \{1,2\}.
}

Using (A.2) and (A.12)
allows us to write
explicit expressions for the matrix elements
to leading order for large $\mu$,
\eqn\flargemu{\eqalign{
F_{m(1)}^+ &= \sqrt{2 \mu \alpha_{(1)}} \alpha_{(2)} (-1)^{m+1},\cr
F_{m(2)}^+ &= \sqrt{2 \mu \alpha_{(2)}} \alpha_{(1)},\cr
F_{m(3)}^+ &\sim
{1 \over \pi \sqrt{2 \mu |\alpha_{(3)}|}} (-1)^m m \sin (m \pi \beta).
}}
Finally the relation \fminus\ tells us that to leading order
we have
\eqn\prlarge{
F_{(r) n}^- =  i {n \over 2 \mu \alpha_{(r)}}
F_{(r) n}^+, \qquad r\in \{1,2\}, \qquad
\qquad
F_{(3) n}^- = - i {2 \mu \alpha_{(3)} \over n}
F_{(3) n}^+.
}

\newsec{The String Theory Amplitudes}

In this section we use the results of the previous section to
calculate the matrix elements \want\ from
string theory for large $\mu$.
In \SpradlinAR\ the cubic interaction in
the light-cone string field theory Hamiltonian
was determined up to an overall function
$f(\mu p^+ \alpha')$.  
A careful analysis of supergravity scattering in the pp-wave
background \KiemXN\ suggests that there the corresponding
function is a numerical constant.
It is natural to conjecture that this result holds
for the full string theory, although an honest
calculation would require checking that the bosonic
and fermionic functional determinants in \SpradlinAR\ cancel
(as one might expect
from supersymmetry).
We will use the symbol $\sim$ in this section to
indicate this assumption about the overall
normalization.

Since the polarization of all excitations lies in the first SO(4),
the analysis at the end of section 2 shows that for these matrix
elements, the prefactor may be taken to be
\eqn\aaa{
\widehat{H} = \delta^{ij} ( K_+^i + K_-^i)(K_+^j - K_-^j),
\qquad i,j\in \{1,\ldots,4\}.
}
Using this and the expressions \vertex\ and \kpmexpan,
it is simple to show that the desired matrix elements are
\eqn\answers{
\eqalign{
\langle O^{J}_n | H |O^{J_1}_m \rangle |O^{J_2}\rangle
&\sim 2 \pi g_s \mu^2  (F^+_{m(1)} F^+_{n(3)}
- F^-_{m(1)} F^-_{n(3)})(\overline{N}^{(13)}_{m,n}
- \overline{N}^{(13)}_{-m,-n}),\cr
\langle O_n^{J} | H | O_\phi^{J_1} \rangle | O_\psi^{J_2} \rangle
&\sim
2 \pi g_s \mu^2   F_{n(3)} ( F_{0(1)} \overline{N}^{(23)}_{0,n}
+ F_{0(2)} \overline{N}^{(13)}_{0,n})
}}
(written in the $a$ oscillator basis---see
footnote 2).
We stress that the results \answers\ are correct for all $\mu$.
In order to compare with \want\ for large $\mu$, we use the
fact that at $\mu = \infty$ the only
nonzero Neumann matrices are
\eqn\aaa{
\overline{N}^{(r3)} = - \sqrt{ \alpha_{(r)}
\over|\alpha_{(3)}|} X^{(r) {\rm T}}, \qquad r \in \{1,2\}
}
(and their transposes
$\overline{N}^{(3r)}$).
Using the expressions from appendix A, as well as the
results \flargemu\ and \prlarge, we immediately
obtain
\eqn\matrixelem{
\eqalign{
\langle O^{J}_n | H |O^{J_1}_m \rangle |O^{J_2}\rangle&\sim
{4 g_s \mu \over  \pi} (1+\beta)
 \sin(n \pi  \beta)^2,
\cr
\langle O_n^{J} | H | O_\phi^{J_1} \rangle | O_\psi^{J_2} \rangle
&\sim
{4  g_s \mu \over  \pi}
\sqrt{-\beta(1+\beta)}
\sin(n \pi \beta)^2.
}}
The second matrix element agrees precisely with \want\ after
recalling that $\beta = - y$, but the first matrix element
disagrees.

Using the matrix elements of the prefactor calculated in
this paper, it is easy to comment on more general matrix elements
which have not yet been studied in the gauge theory.
Already in section 2 we mentioned that the tensor structure
of the prefactor is such that we expect the matrix
elements to change sign when $\phi$ and $\psi$ impurities
are all changed to $D_\mu Z$'s,
while we expect vanishing
matrix elements when one string has only $\phi$ or $\psi$
impurities and the other has only $D_\mu Z$ impurities.
Another class of amplitudes which were not studied in \seven\ are
those in which the total number of impurities is not conserved
in the interaction.
From the string theory point of view these interactions
are not qualitatively different from the ones in which the
number of impurities is conserved, and the matrix elements
are easily calculated using the formulas from this paper.
For large $\mu$ they typically scale as
\eqn\predic{\eqalign{
2~{\rm impurities} + 0~{\rm impurities} \to
2~{\rm impurities}&=
{\cal{O}}(\mu),\cr
1~{\rm impurity} + 1~{\rm impurity} \to
2~{\rm impurities}&=
{\cal{O}}(\mu),\cr
2~{\rm impurities} + 0~{\rm impurities} \to
0~{\rm impurities}&=
{\cal{O}}(\mu^2),\cr
0~{\rm impurities} + 0~{\rm impurities} \to
2~{\rm impurities} &=
{\cal{O}}(\mu^2).
}}
Finally,
a class of matrix elements which
has not yet been studied on either side
of the correspondence
(see however \fgm)
are those which involve non-primary
string states (i.e., states with fermionic oscillators).
On the string theory side, one would need to know the matrix
elements of the fermionic operator $\Lambda$.

\smallskip

\noindent
{\bf Note Added.}
The original version of this paper reported agreement between the
matrix elements \matrixelem\ calculated in string field theory and those
predicted by the proposal \conjecture.
The source of the spurious agreement stemmed from a missing
$i$ in the quantity $s(p)$ introduced in \yyy, such that
the $i$ was missing from \fminus.  As a consequence,
the $F_{m(1)}^- F_{n(3)}^-$ term in \answers\ was reported
with the wrong sign.
As mentioned in the introduction, the resulting apparent disagreement
between light-cone string field theory and the proposal \conjecture\ in
no way disparages the BMN correspondence, but merely indicates that
(as mentioned under \dictionary)
one should take into account the fact that at finite $g_2$,
the dictionary between single- (double-, etc.) string states and
single- (double-, etc.) trace BMN operators gets corrected
\refs{\basisone \basistwo-\basisthree}.

Thanks to the Internet archive, we are able to go back in time and
set the record straight.  We refer the reader to \psvvv\ for a
clarification of the basis transformation between the gauge theory
and string field theory.
At the end of the day, it turns out that the prefactor for the
cubic Hamiltonian $H_3$ (for string states which are primary
in the notation of subsection 2.2) is proportional to
\eqn\aaa{
\sum_{r=1}^3 \sum_{n=-\infty}^\infty {\omega_{n(r)} \over
\alpha_{(r)}} \alpha_{n(r)}^{I \dagger} \alpha_{-n(r)}^J v_{IJ},
}
where $\alpha_n$  are the oscillators in the BMN basis (see footnote 2)
and
$v_{IJ} = {\rm diag}(1_4,-1_4)$.
This simple formula is valid for all $\mu$.

\bigskip

\centerline {\bf Acknowledgements}

\smallskip

We are grateful to D. Freedman, M. Headrick, I. Klebanov,
S. Minwalla, L. Motl and A. Strominger for useful discussions and comments.
M.S. is supported by DOE grant DE-FG02-91ER40671, and
A.V. is supported by DE-FG02-91ER40654.

\appendix{A}{A Compendium of Matrices and Their Properties}

In this appendix we write down the infinite matrices
which appear in the light-cone string vertex
and list some of their properties.
Some of the matrices are labelled by indices running
from $-\infty$ to $\infty$, while others are labelled by indices
running only from $1$ to $\infty$.

We define $\beta = \alpha_{(1)}/\alpha_{(3)}$.  The delta-functional
overlap involves the matrices
\GreenTC
\eqn\adef{\eqalign{
A^{(1)}_{mn}
&=(-1)^{m+n+1} {2 \sqrt{m n} \over \pi} 
{\beta \sin{m \pi \beta}\over n^2-m^2 \beta^2},
\cr
A^{(2)}_{mn}
&=(-1)^{m+1} {2 \sqrt{m n} \over \pi} 
{(\beta+1) \sin{m \pi \beta} \over n^2-m^2 (\beta + 1)^2},
\cr
A^{(3)}_{mn} &= \delta_{mn},\cr
C_{mn} &= m \delta_{mn},
}}
and the vector
\eqn\bdef{
B_m =
(-1)^{m+1} {2 \over \pi} {\alpha_{(3)} \over
\alpha_{(1)} \alpha_{(2)}} m^{-3/2} \sin m \pi \beta.
}
These are all defined for $m,n>0$ only.
In \GreenTC\ it was shown that the identities
\eqn\idone{\eqalign{
A^{(r) {\rm T}} C^{-1} A^{(s)} = - {\alpha_{(r)}
\over \alpha_{(3)}} \delta^{rs} C^{-1}, \qquad
A^{(r) {\rm T}} C A^{(s)} = - {\alpha_{(3)}
\over \alpha_{(r)}} \delta^{rs} C, \qquad
A^{(r) {\rm T}} C B = 0
}}
hold for $r,s\in \{1,2\}$,
and that
\eqn\idtwo{
\eqalign{
\sum_{r=1}^3 {1 \over \alpha_{(r)}} A^{(r)} C
A^{(r) {\rm T}} &= 0,\cr
\sum_{r=1}^3 \alpha_{(r)} A^{(r)} C^{-1} A^{(r) {\rm T}}
&= \ha \alpha_{(1)}
\alpha_{(2)} \alpha_{(3)} B B^{\rm T}.
}}
Occasionally we will use the shorthand
\eqn\aaa{
A_-^{(r)} \equiv {\alpha_{(3)} \over \alpha_{(r)}}
C^{-1} A^{(r)} C,
}
so that the first identity in \idone\ takes the form
\eqn\aaa{
A_-^{(r) {\rm T}} A^{(s)} = -\delta_{rs} {\bf 1}, \qquad r,s \in \{1,2 \}.
}

In \SpradlinAR\ these matrices were assembled into the larger matrices
$X^{(r)}$, defined for $r=3$ by $X^{(3)} = {\bf 1}$
and for $r,s\in \{1,2\}$ by
\eqn\xmatdef{
\eqalign{
X^{(r)}_{mn} &=
(C^{1/2} A^{(r)} C^{-1/2})_{mn}
\qquad\qquad\qquad{\rm if}~m,n>0,\cr
&= {\alpha_{(3)} \over \alpha_{(r)}} (C^{-1/2} A^{(r)} C^{1/2})_{
-m,-n},
\qquad{\rm if}~m,n<0,\cr
&=  - {1 \over \sqrt{2}} \epsilon^{rs} \alpha_{(s)} (C^{1/2} B)_m\qquad
\qquad~~~{\rm if}~n=0~{\rm and}~m>0,\cr
&=1\qquad\qquad\qquad\qquad\qquad\qquad\qquad{\rm if}~m=n=0,\cr
&= 0\qquad\qquad\qquad\qquad\qquad\qquad\qquad{\rm otherwise}.
}}
From \idone\ it follows that
\eqn\aaa{
X^{(r) {\rm T}} X^{(s)} = - {\alpha_{(3)} \over
\alpha_{(r)}} \delta_{rs} {\bf 1}
}
for $r,s\in \{1,2\}$, while \idtwo\ implies that
\eqn\aaa{
\sum_{r=1}^3 \alpha_{(r)} X^{(r)} X^{(r) {\rm T}} = 0.
}

Next we define the diagonal matrices
\eqn\aaa{
\left[ C_{(r)} \right]_{mn} = \delta_{mn} \omega_{m(r)}
=  \delta_{mn} \sqrt{m^2 + \mu^2 \alpha_{(r)}^2}.
}
We will use these matrices both in formulas where
$m$ ranges over all integers as well as in formulas where
$m$ ranges only over positive integers---in each formula
it should be clear what the intended summation range is.
For example, when we define
\eqn\aaa{
U_{(r)} = C^{-1} (C_{(r)} - \mu \alpha_{(r)} {\bf 1}),
}
the presence of $C^{-1}$ makes it clear that
this is for positive indices only.

Finally, when investigating the large $\mu$
limit of the prefactor, we will need to know
\eqn\aaa{\eqalign{
(A^{(1) {\rm T}} C^{-1} B)_n &= 2 (-1)^n n^{-3/2} {\beta \over
\alpha_{(3)}},\cr
(A^{(2) {\rm T}} C^{-1} B)_n &= 2 n^{-3/2}
{\beta+1 \over \alpha_{(3)}}.
}}
These results
follow easily from the definitions
\adef\ and \bdef\ with the help of the sum
\eqn\coolsum{
\sum_{p=1}^\infty {1 \over p^2}
{\sin^2 (p \pi \beta) \over n^2 - p^2 \beta^2}
= - \ha {\pi^2 \over n^2} \beta(\beta+1), \qquad
-1 < \beta < 0.
}

\appendix{B}{On the Neumann Matrices}

The Neumann matrix elements $\overline{N}^{(rs)}_{mn}$
in the pp-wave background were determined in \SpradlinAR.
Formally, they are given by
\eqn\neumanndef{
\overline{N}^{(rs)} = \delta^{rs} {\bf 1}
- 2 C_{(r)}^{1/2} X^{(r) {\rm T}} \Gamma_a^{-1}
X^{(s)} C_{(s)}^{1/2},
}
where
\eqn\gammaadef{
\Gamma_a = \sum_{r=1}^3 X^{(r)} C_{(r)} X^{(r) {\rm T}}.
}
The formula \neumanndef\ is not convenient for actually
calculating matrix elements because the matrix \gammaadef\ does not exist.
This technical
nuisance arises already in flat space, and can be cured with
the help of a trick which we now explain.

From the structure of the $X$ matrices \xmatdef\ it is clear
that $\Gamma_a$ is block diagonal.  Using the identities \idtwo,
it is easy to see that the three blocks can be written as
\eqn\gammasplit{
\left[ \Gamma_a \right]_{mn} =
\cases{
(C^{1/2} \Gamma_+ C^{1/2})_{mn}& $m,n>0$,\cr
-2 \mu \alpha_{(3)}& $m=n=0$,\cr
(C^{1/2} \Gamma_- C^{1/2})_{-m,-n} & $m,n<0$,}
}
where we have defined
\eqn\gpmdef{\eqalign{
\Gamma_+ &= \sum_{r=1}^3 A^{(r)} U_{(r)}  A^{(r) {\rm T}},\cr
\Gamma_- &= \sum_{r=1}^3
 A_-^{(r)}
U_{(r)}^{-1} A^{(r) {\rm T}}_-,
}}
and the factors of $C^{1/2}$ are included for convenience
in \gammasplit\ so that $\Gamma_+$ reduces to the matrix
$\Gamma$
of \GreenTC\ when $\mu$ is set to zero.
The matrix $\Gamma_+$ exists and is invertible, but
the matrix $\Gamma_-$ does not exist because the matrix
product in \gpmdef\ leads to a divergent sum.
Fortunately, it is $\Gamma_-^{-1}$, not $\Gamma_-$,
which appears in the
Neumann matrices.  
In order to define $\Gamma_-^{-1}$, one can use the identities
\idone\ to prove (formally) that
\eqn\aaa{
\Gamma_+ \Gamma_- = \Gamma_+ U_{(3)}^{-1}
+ U_{(3)} \Gamma_-.
}
Multiplying by $\Gamma_+^{-1}$ from the left
and $\Gamma_-^{-1}$ from the right leads to an identity which
allows us to {\bf define} $\Gamma_-^{-1}$:
\eqn\gmdef{
\Gamma_-^{-1} = U_{(3)} - U_{(3)} \Gamma_+^{-1} U_{(3)}.
}
This matrix $\Gamma_-^{-1}$ is well-defined but not invertible,
despite the unfortunately misleading notation.

Using the definition \gpmdef\ and the identities \idone, it 
is easy to show that
\eqn\aaa{
\Gamma_+^{-1} C^{-1} U_{(3)} A^{(r)} =
C^{-1} A_{(r)} + {\alpha_{(r)}\over\alpha_{(3)}}
\Gamma_+^{-1} A_{(r)} C^{-1} U_{(3)}, \qquad r \in \{1,2\}.
}
Then, using the definition \gmdef, it follows that
\eqn\aaa{
\Gamma_-^{-1} A_-^{(r)} = - U_{(3)} \Gamma_+^{-1} A_{(r)} U_{(r)}, \qquad
r\in \{1,2\}.
}
This remarkable identity allows us to write down
simple relations between Neumann matrix elements
for positive $m,n$ indices and negative $m,n$ indices.
For example, for $r \in \{1,2\}$ and $m,n>0$ we have
\eqn\aaa{
\left[ \overline{N}^{(r3)} \right]_{-m,-n} = -
\left[ U_{(r)}
\overline{N}^{(r3)} U_{(3)} \right]_{mn}, \qquad r\in \{1,2\}.
}
Relations for other $r,s$ are obtained similarly.

\appendix{C}{On the Large $\mu$ Expansion of the Matrix $\Gamma_+$}

In this appendix we summarize what is known about
the behavior of the matrix $\Gamma_a$ for large $\mu$.
This question has been studied in \refs{\HuangWF,\ChuPD}.
We demonstrate that matrix elements of the light-cone Hamiltonian
will typically have terms of order $(\lambda')^{3/2}$, which
are puzzling from the field theory point of view.

The matrix $\Gamma_a$ \gammaadef\ encodes much of the 
structure of the cubic string vertex, but it is a very
complicated function of $\mu$.
The large $\mu$ expansion is difficult to study
because the process of taking
$\mu \to \infty$ limit does not commute
with the infinite sums in the matrix multiplication which
defines $\Gamma_a$.
A prototype of this phenomenon is the sum
\eqn\knownsum{
\sum_{p=1}^\infty
{1 \over p^2} {1 \over 1 + \lambda p^2} = {\pi^2 \over 6}
- {\pi \over 2} \sqrt{\lambda}  + {\cal{O}}(\lambda),
}
where we have expanded the known result for small $\lambda$.
Naively expanding the summand would lead one to the
incorrect
conclusion
that
only integer powers of $\lambda$ appear.
Of course in this toy example we know how to do the sum \knownsum\ explicitly.
For $\Gamma_a$ the sums are much more complicated, but we
will be able to show that a similar phenomenon occurs.

We start by recalling that
\eqn\gpone{\eqalign{
\Gamma_+ &= \sum_{r=1}^3 A^{(r)} U_{(r)}
 A^{(r) {\rm T}}\cr
&= U_{(3)} +
\sum_{r=1}^2 A^{(r)}  U_{(r)} A^{(r) {\rm T}}.
}}
If we split off the leading term for large $\mu$, which is
allowed since it gives a convergent sum, we obtain
\eqn\obtain{\eqalign{
\Gamma_+ &= U_{(3)} + {1 \over 2 \mu} \sum_{r=1}^2
{1 \over \alpha_{(r)}} A^{(r)} C A^{(r) {\rm T}}
+ \sum_{r=1}^2 A^{(r)} \left(
U_{(r)} - {C \over 2 \mu \alpha_{(r)}} \right) A^{(r) {\rm T}}
\cr
&=
U_{(3)} - {1 \over 2 \mu \alpha_{(3)}} C + R,
}}
where we have used \idtwo\ to perform one
of the summations, and defined
the `remainder' matrix $R$ which has the matrix elements
\eqn\rdef{
\eqalign{
R_{mn} &=  \sum_{r=1}^2
\left[
A^{(r)} \left( U_{(r)} - {C \over 2 \mu \alpha_{(r)}} \right) A^{(r) {\rm T}}
\right]_{mn}\cr
&= {4 \over \pi^2} (-1)^{m+n}
\sqrt{m n} \sin (m \pi \beta) \sin(n \pi \beta)
\left( \Sigma_{mn}(\beta) + \Sigma_{mn}(-1-\beta)\right),
}}
written in terms of the sum
\eqn\aaa{
\Sigma_{mn}(\beta) = \beta^2 \sum_{p=1}^\infty
{p \over (p^2 - m^2 \beta^2)(p^2 - n^2 \beta^2)}
\left[
{ \sqrt{p^2 + \mu^2 \alpha_{(3)}^2 \beta^2} - \mu \alpha_{(3)}
\beta
\over p} - {p \over 2 \mu \alpha_{(3)} \beta} \right].
}
Naively taking
$\mu$ to infinity before performing the sum would give
$\mu^{-3}$ times a linearly divergent sum.
Instead, it is clear that one must perform the sum first
at finite $\mu$ (where it is clearly convergent,
since the term in brackets behaves like $p$ for
large $p$) and then
take the large $\mu$ limit of the result.
It is easy to estimate that the resulting behavior
is of order $\mu^{-2}$, and that there is no cancellation
between the two terms in \rdef, so that the matrix
$R$ has a nonzero leading term of order $\mu^{-2}$.

Going back to \obtain, we are able to conclude that for large
$\mu$,
\eqn\endform{
\Gamma_+ = - {2 \mu \alpha_{(3)}} C^{-1}
- {1 \over \alpha_{(3)} \mu} C
+ {\cal{O}}(\mu^{-2}),
}
where we have proven that the $\mu^{-2}$ term is nonzero,
but we have not calculated it explicitly.
The authors of \ChuPD\ used an analytic continuation
argument to sum all of the odd powers of $\mu$ on the
right-hand side of \endform, obtaining
the remarkably simple result
\eqn\aaa{
\Gamma_+ = 2 C_{(3)} C^{-1}.
}
Additional terms must be present
since
it is manifest from \gpone\ that
$\Gamma_+$ 
goes over smoothly 
to the flat
space matrix $\Gamma$
from \GSB\ (which has a complicated
structure) as $\mu \to 0$.
Certainly, it would be very desirable to have better
analytic control over the matrix $\Gamma_+$ (or
even better, $\Gamma_+^{-1}$).

The presence of the $\mu^{-2}$ term in
\endform\ implies the existence of $(\lambda')^{3/2}$ terms
in typical light-cone Hamiltonian
matrix elements, which seem difficult to explain from
the gauge theory.
Perhaps the prefactor contains half-integer powers of
$\lambda'$ in such a way that these strange powers
cancel, or perhaps the dictionary \conjecture\ must
be modified in an appropriate way at higher
order to achieve agreement.

\listrefs

\end